\pdfoutput=1
\documentclass[%
aps,
reprint,
superscriptaddress,
]{revtex4-1}

\usepackage{graphicx}
\usepackage{dcolumn}
\usepackage{bm}
\usepackage{epstopdf}
\usepackage{xr}
\usepackage{amsmath}
\usepackage{color}
\usepackage[normalem]{ulem}
\usepackage{wasysym}
\usepackage{siunitx}
\usepackage{physics}
\usepackage{lineno}
\usepackage[T1]{fontenc}
\usepackage{booktabs}
\usepackage[section]{placeins}
\usepackage{textcomp}
\usepackage{amssymb}
\usepackage[english]{babel}

\usepackage{hyperref}
\hypersetup{colorlinks=true,%
citecolor=blue,%
filecolor=black,%
linkcolor=black,%
urlcolor=blue
}

\makeatletter

\newcommand{\Rmnum}[1]{\expandafter\@slowromancap\romannumeral #1@}
\makeatother

\begin{document}

\markboth{\jobname}{\jobname .tex}

\title{On-demand local modification of high-$T_\text{c}$ superconductivity in few unit-cell thick Bi$_2$Sr$_2$CaCu$_2$O$_{8+\delta}$}
\author{Sanat Ghosh}
\affiliation{Department of Condensed Matter Physics and Materials Science, Tata Institute of Fundamental Research, Homi Bhabha Road, Mumbai 400005, India}

\author{Jaykumar Vaidya}
\affiliation{Department of Condensed Matter Physics and Materials Science, Tata Institute of Fundamental Research, Homi Bhabha Road, Mumbai 400005, India}

\author{Sawani Datta}
\affiliation{Department of Condensed Matter Physics and Materials Science, Tata Institute of Fundamental Research, Homi Bhabha Road, Mumbai 400005, India}

\author{Ram Prakash Pandeya}
\affiliation{Department of Condensed Matter Physics and Materials Science, Tata Institute of Fundamental Research, Homi Bhabha Road, Mumbai 400005, India}

\author{Digambar A. Jangade}
\affiliation{Department of Condensed Matter Physics and Materials Science, Tata Institute of Fundamental Research, Homi Bhabha Road, Mumbai 400005, India}

\author{Ruta N. Kulkarni}
\affiliation{Department of Condensed Matter Physics and Materials Science, Tata Institute of Fundamental Research, Homi Bhabha Road, Mumbai 400005, India}

\author{Kalobaran Maiti}
\affiliation{Department of Condensed Matter Physics and Materials Science, Tata Institute of Fundamental Research, Homi Bhabha Road, Mumbai 400005, India}

\author{A. Thamizhavel}
\affiliation{Department of Condensed Matter Physics and Materials Science, Tata Institute of Fundamental Research, Homi Bhabha Road, Mumbai 400005, India}

\author{Mandar M. Deshmukh}
\homepage{deshmukh@tifr.res.in}
\affiliation{Department of Condensed Matter Physics and Materials Science, Tata Institute of Fundamental Research, Homi Bhabha Road, Mumbai 400005, India}

\begin{abstract}

High-temperature superconductors (HTS) are important for potential applications and for understanding the origin of strong correlations. Bi$_2$Sr$_2$CaCu$_2$O$_{8+\delta}$ (BSCCO), a van der Waals material, offers a platform to probe the physics down to a unit-cell. Guiding the flow of electrons by patterning 2DEGS and oxide heterostructures has brought new functionality and access to new science. Similarly, modifying superconductivity in HTS locally, on a small length scale, will be of immense interest for superconducting electronics. Here we report transport studies on few unit-cell thick BSCCO and modify its superconductivity locally by depositing metal on the surface. Deposition of chromium (Cr) on the surface over a selected area of BSCCO results in insulating behavior of the patterned region. Cr locally depletes oxygen in CuO$_2$ planes and disrupts the superconductivity in the layers below. Our technique of modifying superconductivity is suitable for making sub-micron superconducting wires and more complex superconducting electronic devices.

\end{abstract}

\maketitle


High-temperature superconductors (HTS) are rich systems to explore correlation physics although the complete microscopic description of its superconductivity is still missing \cite{lee_doping_2006}. Tuning of these correlations with external controls provide insights about the nature of the interactions and is interesting for fundamental science and applications. Doping, in general, is one such control. One particular method is surface doping and can be accessed by placing impurity atoms on the surface of HTS. Spectroscopic studies on such HTS systems in 3D reveal that the impurities only affect the layers very close to the surface while the bulk remains unaffected \cite{luo_mechanisms_1992}.
Of various members of the HTS family, BSCCO \cite{maeda_new_1988} is one of the most studied compounds as it can be easily cleaved due to its van der Waals structure to reveal pristine surface for spectroscopic surface studies.

Studies on 2D materials \cite{novoselov_two-dimensional_2005} over the years  have led to exciting new insights into their electronic and optical properties. The cleavable nature of BSCCO has opened up possibilities of realizing high-$T_\text{c}$ superconductivity in 2D limit. Initial investigations found that one unit-cell thick BSCCO is insulating \cite{novoselov_two-dimensional_2005} due to the distortion of the oxygen stoichiometry; this challenge was overcome for the first time by covering few unit-cell BSCCO with graphene \cite{jiang_high-_2014}. Recently it has been shown that the superconducting critical temperature, $T_\text{c}$ can be modulated in few unit-cell thick BSCCO by modulating density of charge carriers with electrostatic gating \cite{sterpetti_comprehensive_2017,liao_superconductorinsulator_2018}. Detailed study of phase diagram and spectroscopic analysis, using ozone induced doping, has been explored for monolayer BSCCO \cite{yu_high-temperature_2019}. Experiment probing the possible role of superconducting fluctuations has also been carried out on two unit-cell thick BSCCO \cite{zhao_sign-reversing_2019}.

For studying and modulating electronic properties, it is desirable to pattern the system to guide the flow of electrons spatially. Semiconductor heterostructure based two-dimensional electron gas (2DEG) systems are one of the most celebrated examples where the electron gas can be controlled by patterning with electrostatic gates. For the case of oxide nanoelectronics \cite{cen_oxide_2009}, similar advances have been made, leading to more in-depth insight \cite{cheng_electron_2015} about the microscopic nature of electronic interactions. Devices made by patterning of YBa$_2$Cu$_3$O$_{7 + \delta}$ using helium ions provide one possible strategy for realizing superconducting quantum interference devices \cite{cybart_nano_2015}. For many device applications, these existing methods have limitations as they are inefficient in modulating superconducting $T_\text{c}$ locally, and ion bombardment can lead to structural damages. A rational and scalable process for tuning $T_\text{c}$ locally and patterning on small length scale is of potential interest.


In this letter, we report a study where we alter superconductivity locally on $\sim$ \SI{1}{\micro\metre} length scale in few unit-cell thick BSCCO by exploiting van der Waals nature of the system. Deposition of a thin pattern of Cr, and Ti, on top of few unit cell thick BSCCO flake, modifies the superconductivity of BSCCO underneath -- to an insulator, or to a superconductor with lower $T_\text{c}$ respectively. We attribute this alteration of superconductivity in BSCCO to the depletion of oxygen stoichiometry by surface deposition of reactive metals. As a control, we perform similar experiments replacing chromium pattern with that of gold, and find superconductivity to be unaltered supporting our conclusion. We also provide direct evidence using X-ray photoemission spectroscopy (XPS) of Cr altering the structure of CuO$_2$ planes which are responsible for superconductivity in cuprate superconductors.  We further demonstrate the potential of this technique by realizing a device of narrow BSCCO wire geometry. 

\begin{figure*}
\includegraphics[width=15.5cm]{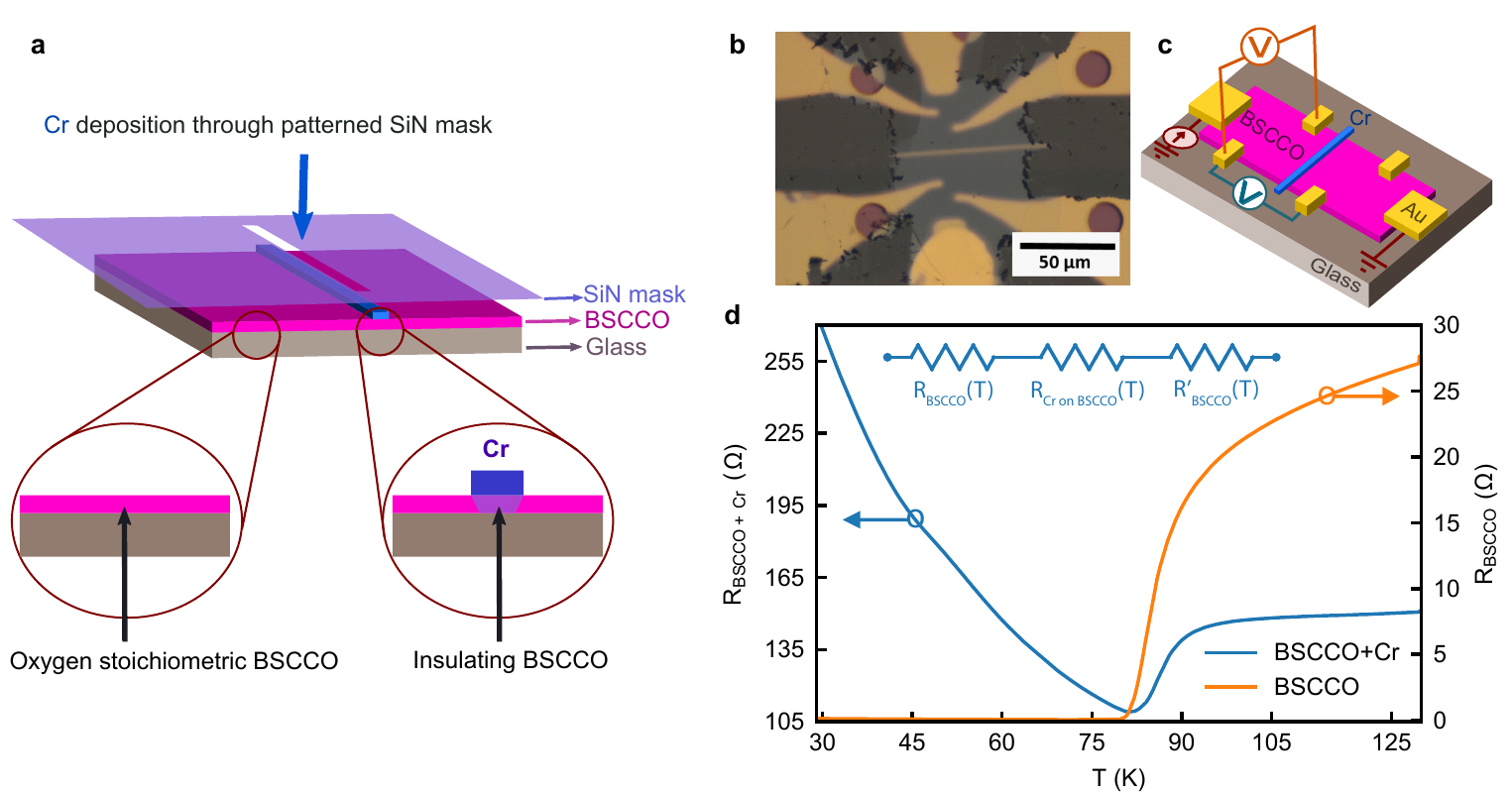}
\caption{ \label{fig:fig1} \textbf{Tuning superconductivity spatially by surface metal deposition using a stencil mask.} (a) Schematic for the deposition of metal film on an exfoliated flake of BSCCO through a patterned SiN mask. 2D BSCCO flakes of $\sim$ \SI{100}{\micro\metre} lateral dimension are exfoliated by anodic bonding technique \cite{balan_anodic_2010,sterpetti_comprehensive_2017} b) Optical image of the device after shaping the BSCCO flake. We deposit \SI{1}{\micro\metre} wide and 15~nm thick Cr line to modify superconductivity of BSCCO underneath followed by 10~nm Au to protect Cr from oxidation due to ambient. The electrodes that contact BSCCO flake for measurement are made using gold; no sticking layer was used. (c) Schematic of the device architecture and measurement scheme. The orange-colored voltmeter measures the electrical response of pristine BSCCO, while blue voltmeter probes the response across the region of BSCCO patterned locally with Cr deposition. (d) $R$ vs. $T$ data across BSCCO along with the region involving Cr line. Resistance across BSCCO goes to zero below $T_\text{c}$, while resistance across Cr drops to a value of $\sim$ 105~$\Omega$ at $T_\text{c}$ of BSCCO and shows insulating, $\frac{dR}{dT}<0$, behavior below it. Inset shows a simple series resistor model for the region of BSCCO underneath Cr and adjacent two pristine BSCCO wings.}
\end{figure*}


We exfoliate thin flakes of BSCCO from bulk crystals by anodic bonding method \cite{balan_anodic_2010,sterpetti_comprehensive_2017} on borosilicate glass. We prefer this over scotch tape technique as it gives areas of uniform thicknesses as large as $\sim$ \SI{100}{\micro\metre} $\times$ \SI{100}{\micro\metre}. We avoid any chemical processing in fabricating devices as it degrades the quality of BSCCO. Contacts to the flake are thus made by direct metal deposition through a patterned SiN mask \cite{deshmukh_nanofabrication_1999} as shown schematically in Fig.~\ref{fig:fig1}a. The entire process of device fabrication involves two steps of evaporation. In the first step, we deposit 70~nm of gold electrodes to make electrical contacts to BSCCO. Second step involves deposition of 15~nm thick and \SI{1}{\micro\metre} wide Cr line, or any other desired pattern, over selected area of BSCCO for altering its superconductivity. We then deposit 10~nm of Au on top in the same step to protect the Cr line from being oxidized. Electrical measurements are done using low-frequency ac lock-in detection technique in four-probe configuration (for more details, see Methods).


\begin{figure*}
\includegraphics[width=15.5cm]{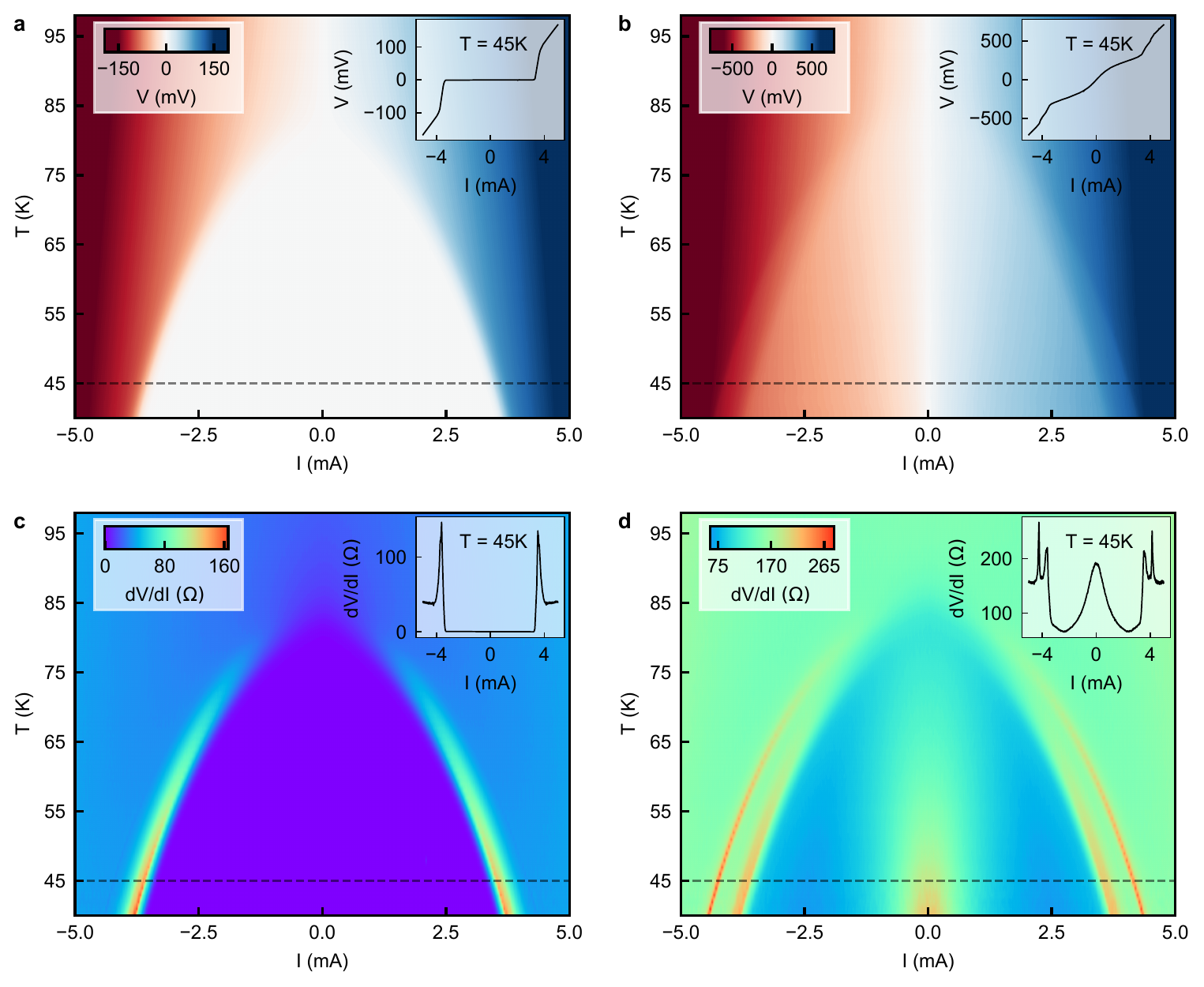}
\caption{ \label{fig:fig2} \textbf{Current--voltage characteristics of a superconductor-insulator-superconductor device made using BSCCO.} (a) Color scale plot of dc voltage ($V$) drop across BSCCO as a function of dc biasing current ($I$) and temperature ($T$). White shaded dome region is the superconducting phase where there is no voltage drop below a certain critical current $I_\text{c}$ and below superconducting critical temperature $T_\text{c}$. Inset shows the current-voltage characteristic across BSCCO at 45~K, a line cut taken along the dotted line indicated in the color scale plot. (b) Color scale plot of dc $I$-$V$ characteristics as a function of temperature across Cr line patterned region. It does not have a dissipationless superconducting region as the resistance across Cr line does not go to zero as shown in Fig.~\ref{fig:fig1}d. Inset shows $I$-$V$ characteristic at 45~K. (c) Color scale plot of differential resistance across BSCCO as a function of dc current and temperature, measured by adding a small ac current with the applied dc current. (d) Color scale plot of differential resistance across Cr line as a function of dc current and temperature. Inset shows a line cut at 45~K indicated by a dashed line on the color scale plot. We see a zero bias resistance peak across the Cr region, as shown in the inset plot.}
\end{figure*}

Fig.~\ref{fig:fig1}b shows an optical image of our device with a pattern of Cr line in the middle. We design our device with multiple electrodes such that in a single device we can study the effect of depositing Cr and compare it with the region of pristine BSCCO that does not have any exposure to Cr; the measurement on multiple probes is done simultaneously. A schematic of the measurement scheme is shown in Fig.~\ref{fig:fig1}c. The device is designed so that we current bias the extremal electrodes and simultaneously measure pairwise resistance of region with only BSCCO and region that includes Cr deposited on BSCCO; this serves as an important control for ensuring that BSCCO remains stoichiometric. Fig.~\ref{fig:fig1}d shows resistance as a function of temperature ($R$ vs. $T$) for region covered by Cr and the pristine BSCCO. The resistance across BSCCO  drops to zero while the resistance across Cr deposited line drops down to $\sim$ 105~$\Omega$ at the transition temperature of BSCCO rather than going to zero. Moreover, below $T_\text{c}$ of BSCCO, the region with Cr on surface shows insulating behavior.

To understand the physical response, we use a series resistor model as a function of temperature for the region, including the Cr covered BSCCO. The model consists of three resistors -- two pristine BSCCO segments, that form the ``wings'', and the BSCCO underneath Cr, which forms the central part of the device response as schematically shown in the inset of Fig.~\ref{fig:fig1}d. The overall nature of $R$ vs. $T$  curve in the normal state of BSCCO is resultant of these two components. The drop in resistance at transition temperature corresponds to the transition of pristine BSCCO to zero resistance state. Hence, below the transition temperature, the response is entirely due to the part of BSCCO underneath Cr. The $R$ vs. $T$ curve across Cr patterned line below $T_\text{c}$ suggests that the deposition of Cr makes the BSCCO underneath insulating. We fit the segment of the curve below $T_\text{c}$ with the Arrhenius model and estimated the energy gap to be 12.44~meV. Moreover, we find that the barrier energy increases as the width of the deposited Cr line is increased (see Supplementary Section 2).


An additional way to characterize the superconductivity is the unique current-voltage ($I$-$V$) response. Fig.~\ref{fig:fig2}a shows a color scale plot of dc $I$-$V$ characteristics of pristine BSCCO as a function of temperature. The white shaded dome area is the dissipationless superconducting transport regime where no voltage drop appears below a critical temperature ($T_\text{c}$) and a critical current ($I_\text{c}$). As a representative curve, $I$-$V$ characteristic at a fixed temperature (45~K) is shown in the inset of Fig.~\ref{fig:fig2}a. Fig.~\ref{fig:fig2}b shows simultaneously measured $I$-$V$ characteristics as a function of temperature across Cr covered BSCCO region. The dome-shaped region in this case is not dissipationless below $T_\text{c}$ as there is finite voltage drop which corresponds to the non-zero finite resistance of region of BSCCO covered with Cr (Fig.~\ref{fig:fig1}d). Boundary of the dome shape region, however, corresponds to the transition of BSCCO to superconducting state which is not covered with Cr and forms the ``wings'' in the series resistor model. In the inset of Fig.~\ref{fig:fig2}b, $I$-$V$ characteristic, at 45~K, across Cr covered BSCCO region is shown as a line cut of the color scale plot. From the inset curve, we see that below $T_\text{c}$, the region of BSCCO, covered with Cr, shows a nonlinear $I$-$V$ response.

\begin{figure}
\includegraphics[width=8cm]{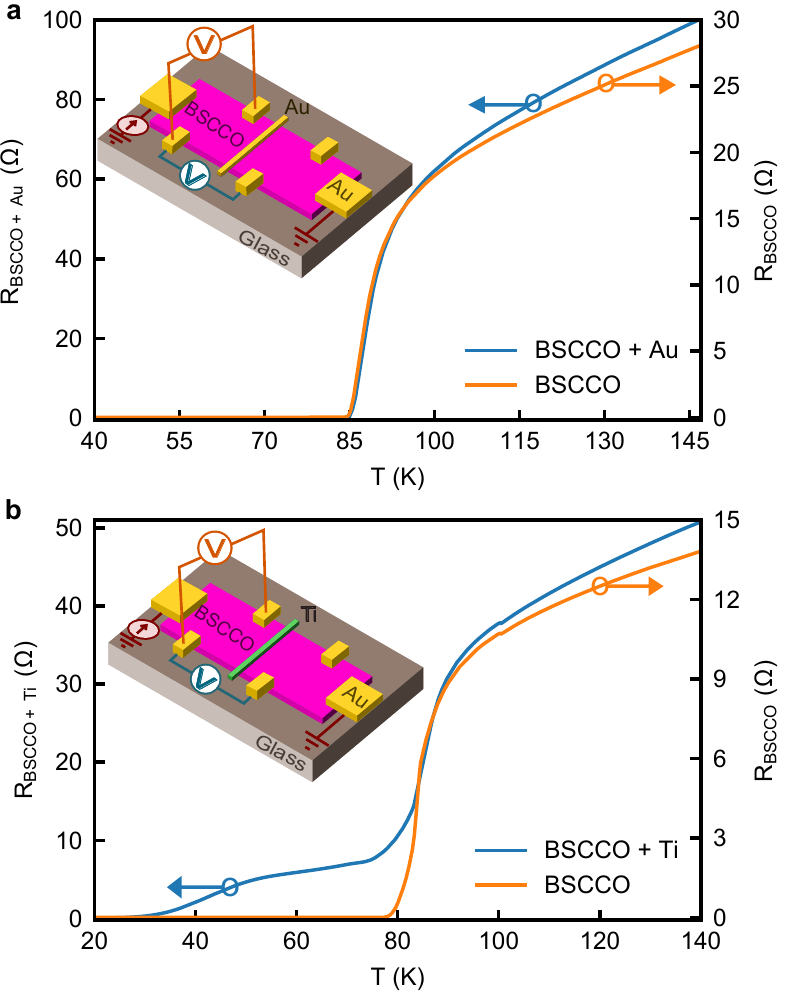}
\caption{ \label{fig:fig3}  \textbf{Role of chemical reactivity in modifying superconductivity.} (a) $R$ vs. $T$ data for the Au line deposited device. The resistance across BSCCO and the Au deposited region drops to zero at the same critical temperature ($T_\text{c}$), suggesting that Au does not affect BSCCO underneath. Inset shows the circuit schematic for this control experiment. (b) $R$ vs. $T$ data of the Ti line deposited device. While the resistance across BSCCO shows usual drop at $T_\text{c}$, the resistance across Ti goes to zero at much lower temperature $\approx$ 25~K. Inset shows the circuit schematic for this control experiment.}
\end{figure}

Fig.~\ref{fig:fig2}c, and Fig.~\ref{fig:fig2}d shows color scale plot of differential resistance ($\frac{dV}{dI}$) vs. dc biasing current ($I$) as a function of temperature across BSCCO and Cr covered BSCCO region, respectively. The respective insets show $\frac{dV}{dI}$ vs. $I$ at 45~K. For the region of BSCCO covered by Cr line, we see a differential resistance peak at zero bias current. The $I$-$V$ and $\frac{dV}{dI}$ response unequivocally suggest that just the surface deposition of Cr is fundamentally modifying the superconducting property of BSCCO underneath.


\begin{figure}
\includegraphics[width=8cm]{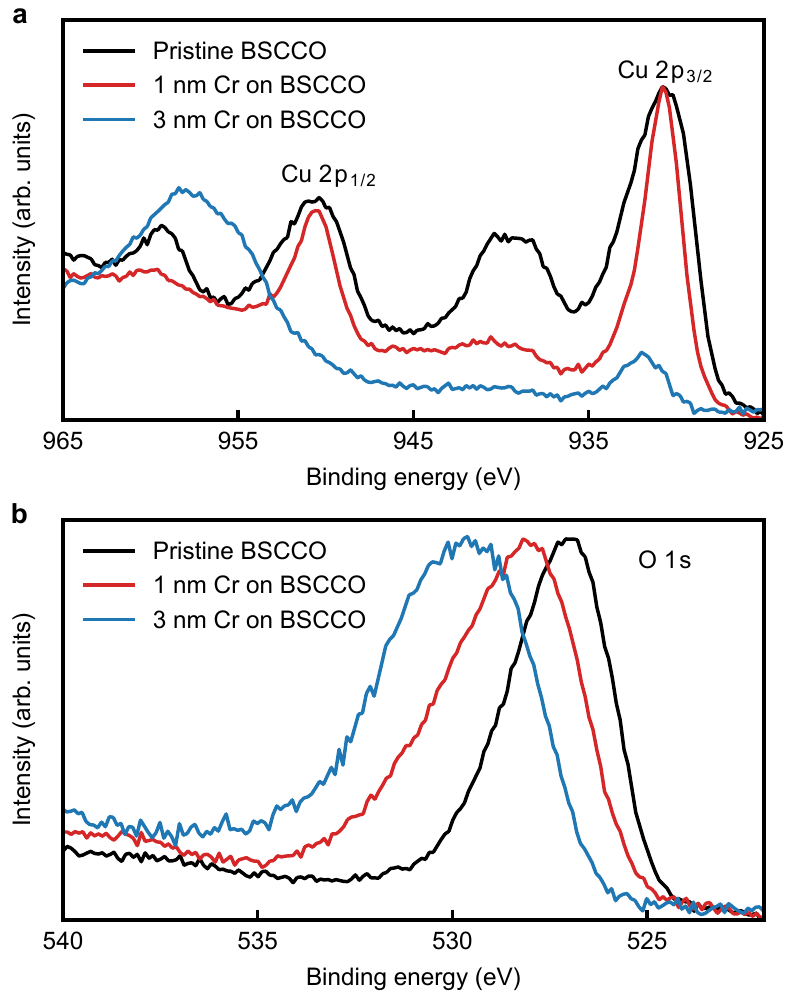}
\caption{ \label{fig:fig4}  \textbf{XPS revealing Cr affects CuO$_2$ planes in BSCCO.} (a) Evolution of Cu 2\textit{p} core electron spectra with increasing thickness of deposited Cr. The peak just next to Cu 2\textit{p$_{3/2}$} is the satellite peak of Cu 2\textit{p$_{3/2}$} and similarly the satellite peak of Cu 2\textit{p$_{1/2}$} is on left side of Cu 2\textit{p$_{1/2}$}. With increasing thickness of Cr, Cu 2\textit{p$_{3/2}$} and Cu 2\textit{p$_{1/2}$} peak becomes narrower than the pristine BSCCO Cu 2\textit{p} peaks. Intensity ratio of main peaks to their satellite peaks also reduces suggesting reduction of Cu$^{2+}$ to Cu$^{1+}$ by removing oxygen from CuO$_2$ layers. (b) Evolution of O 1\textit{s} core electron spectra of bulk BSCCO crystal with thickness of deposited Cr on it. Binding energy of O 1\textit{s} electron shifts towards the higher energy as the thickness of deposited Cr is increased, suggesting a reduction of hole concentration in BSCCO. }
\end{figure}

There can be several possible explanations for our main observation that Cr deposited on the top disrupts the superconductivity in BSCCO below it. One such possible mechanism that could result in this response is the physical damage to BSCCO underneath due to the deposition of energetic atomic clusters during evaporation. If physical damage is the primary mechanism for disrupting BSCCO superconductivity, then it is likely to be independent of the material we deposit by evaporation. To test this hypothesis, we deposit Au instead of Cr. Fig.~\ref{fig:fig3}a shows $R$ vs. $T$ curve across BSCCO along with the region of Au covered BSCCO. Our data suggest that the process of depositing metal by itself does not cause any disruption to BSCCO superconductivity. Therefore, the deposition of Cr disrupts BSCCO superconductivity by a non-physical mechanism, is established from our first control experiment.

Superconductivity in BSCCO and other high $T_\text{c}$ cuprate superconductors arises in the copper oxide planes and $T_\text{c}$ increases with the increasing number of copper oxide planes per unit-cell of the materials \cite{schilling_superconductivity_1993}. In BSCCO, superconductivity critically depends on charge doped into or removed from the CuO$_2$ layers by dopant atoms like oxygen \cite{lee_doping_2006}. On the other hand, Cr is highly reactive and an antiferromagnetic \cite{fawcett_spin-density-wave_1988} metal. So deposition of Cr can affect the BSCCO underneath in two possible ways -- firstly, through its chemical reactivity, it can withdraw oxygen from BSCCO and secondly, being magnetic atom, it can inhibit superconductivity. To disentangle the effect of magnetism from chemical changes that Cr can have on BSCCO, we perform another control experiment by depositing Ti instead. Energetically, the heat of oxide formation of Ti is comparable to that of Cr, as discussed earlier in the context of adatoms on the surface of bulk BSCCO \citep{kimachi_reactive_1991}. Moreover, Ti is a non-magnetic metal. If chemical changes due to reaction are primary mechanisms that disrupt BSCCO superconductivity, depositing Ti should show a similar effect. $R$ vs. $T$ for Ti covered BSCCO device is seen in Fig.~\ref{fig:fig3}b. Strikingly, deposition of Ti shows different behavior from that of Cr -- $T_\text{c}$ across Ti covered BSCCO reduces to 25~K. Moreover, we do not see the insulating nature below $T_\text{c}$, as was the case for Cr deposited device. We have now shown two key results - firstly, with Cr, the BSCCO underneath has an insulating response, and secondly, with Ti deposition the $T_\text{c}$ is modified.


To better understand chemical bonding changes in BSCCO due to Cr deposition, we have used X-ray photoemission spectroscopy (XPS) following past studies that probed the consequence of adatoms on the surface of 3D superconductor \cite{luo_mechanisms_1992}. We use bulk crystals of BSCCO for studying XPS at room temperature (for more experimental details on XPS, see Methods and Supplementary Section 3).

CuO$_2$ planes being responsible for superconductivity in BSCCO, we focus here mainly on evolution of core-level electronic spectra for Cu and O with deposition of Cr. As seen from Fig.~\ref{fig:fig4}a, with the deposition of 1~nm Cr, both Cu 2\textit{p$_{3/2}$} and Cu 2\textit{p$_{1/2}$} peak have become narrower compared to the pristine BSCCO Cu 2\textit{p} peaks. Also, the ratios of intensity of main peaks to their satellite peaks have significantly reduced indicating reduction of Cu$^{2+}$ to Cu$^{1+}$ by removing oxygen from CuO$_2$ planes \citep{weaver_overlayer_1994}. Further deposition of Cr up to 3~nm thickness, the signal from BSCCO is barely visible due to electron escape depth ($\sim$ 1.2~nm) being much smaller than the Cr layer thickness.

Fig.~\ref{fig:fig4}b shows that with the deposition of 1~nm Cr on BSCCO, the O 1\textit{s} peak shifts to higher binding energy due to reduction of hole concentration \citep{harima_chemical_2003}; Fermi level shifts to higher energy, making the binding energy (defined with respect to the Fermi level) higher. For 3~nm deposited Cr, the O 1\textit{s} signal from BSCCO is barely visible, and the primary contribution comes from the diffused oxygen atoms weakly bound to Cr layer and the surface oxygen atoms. These results show that the deposition of Cr on the surface of BSCCO makes the layers underneath deficient in oxygen (which reduces hole concentration significantly), pushing the material towards insulating state.


\begin{figure}
\includegraphics[width=8cm]{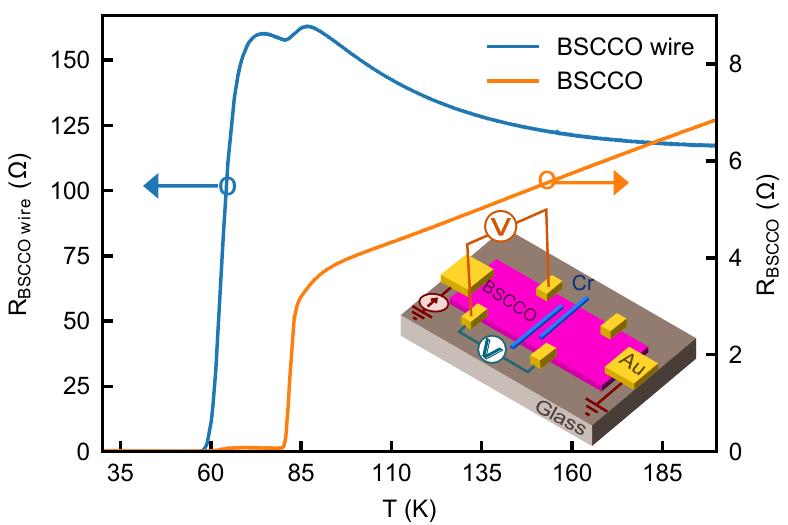}
\caption{ \label{fig:fig5} \textbf{Realizing a narrow BSCCO wire by patterning superconductivity in BSCCO with Cr.} $R$ vs. $T$ curve for a device with $\sim$ 700~nm wide patterned BSCCO wire having $\sim$ \SI{30}{\micro\metre} length shows a distinct insulating $\frac{dR}{dT}<0$ while BSCCO part on the same flake shows metal like response prior to the superconducting transition. The critical temperature for the narrow BSCCO wire is reduced to $\approx$ 57~K. }
\end{figure}

To demonstrate the potential of our technique for patterning 2D BSCCO superconducting devices, we apply our method to realize a narrow channel of BSCCO (for more applications, see Supplementary Section 4). We deposit two lines of Cr separated by a gap which effectively defines a narrow path for supercurrent to flow. The inset of Fig.~\ref{fig:fig5} shows a schematic of the device structure. The length and width of the superconducting channel presented here are $\approx$ \SI{30}{\micro\metre} and $\approx$ 700~nm respectively, suitable for calling it BSCCO wire.  Fig.~\ref{fig:fig5} shows the $R$ vs. $T$ data of the BSCCO wire along with pristine BSCCO region. Resistance across BSCCO wire shows insulating behavior having a drop near $T_\text{c}$ of pristine BSCCO and eventually transitions to superconducting state at $\approx$ 57~K. We also see a reduction of critical current across the patterned BSCCO wire, shown in Supplementary Section 5. Our measurements indicate that it is indeed possible to realize narrow channels of BSCCO which can be used to make devices like single-photon detectors \cite{eisaman_invited_2011} that use an array of the wire geometry that we demonstrate. Single photon detectors working close to liquid N$_2$ temperature are desirable. Also, this geometry can be used to study size effects on superconductivity in high-$T_\text{c}$ materials. 


We have developed a technique to tune superconductivity locally in few unit-cell thick 2D superconductor BSCCO. Upon depositing metal, BSCCO underneath can be turned into an insulator, or a superconductor, with a lower $T_\text{c}$ using Cr and Ti respectively. We have shown the potential device applications of this technique that are currently limited in the smallest dimension by the SiN shadow mask technique \cite{deshmukh_nanofabrication_1999}. Minimum feature size of devices can be further reduced to $\sim$ 50~nm by using $h$-BN shadow masks placed on BSCCO that are covered with metal. Our study could also allow creation of THz emission devices \cite{welp_superconducting_2013} for on-chip THz emission studies of correlated materials. High-$T_\text{c}$ SQUIDs made using our technique could be used for scanned probe studies \citep{vasyukov_scanning_2013} of correlated materials up to higher temperature. While we demonstrate using few unit-cell thick BSCCO, this rationale for tuning superconductivity should also work for other systems based on Cu-O and Ni-O planes \cite{li_superconductivity_2019}.

\section*{Acknowledgments:}

We thank  Abhay Shukla, Arup Kumar Raychaudhuri, Srinivasan Ramakrishnan, and Vikram Tripathi for helpful discussions. We also thank L.D. Varma Sangani for assistance with shaping BSCCO flakes. We acknowledge Department of Science and Technology (DST) Nanomission grant SR/NM/NS-45/2016 and Department of Atomic Energy (DAE) of Government of India for support.

\section*{Author contributions:}
S.G. fabricated the devices and did the measurements. J.V assisted in fabrication. S.G. analyzed the data. S.D., R.P.P. and K.M did XPS measurements and carried out analysis. A.T. grew the BSCCO crystals with D.A.J. and R.N.K. S.G. and M.M.D. wrote the manuscript. All authors provided inputs to the manuscript. M.M.D. supervised the project.


%

\onecolumngrid

\section*{Methods}

\subsection*{Device fabrication}
Few unit-cell thick BSCCO is exfoliated from bulk crystal by anodic bonding method \citep{balan_anodic_2010,sterpetti_comprehensive_2017} on borosilicate glass substrate in ambient. The borosilicate glass is heated at 180~C. A few mm size BSCCO crystal is then placed on the glass. We then apply a high voltage ($\sim$ 900~V) between the crystal and the bottom of the glass substrate for about 10~min. In this process first few layers of BSCCO which are in contact to substrate get electrostatically bonded. After this we remove the remaining bulk crystal by scotch tape. Multiple times peeling can be done to reduce the thickness of the exfoliated flake. 70~nm thick Au contacts to the flake is made by e-beam evaporation through a SiN mask. A thin pattern of Cr line is then deposited in the second evaporation step through a separate SiN mask. 10~nm of Au capping layer is also deposited in the same step to protect Cr from being oxidized. Finally the contacts to the BSCCO flake is extended by depositing Cr/Au (5~nm/70~nm) through a metal mask. We then shape the flake by removing unwanted parts (see Supplementary Section 1) and immediately cooldown the device in a closed cycle cryostat for electrical measurements. 

\subsection*{XPS}
We use bulk crystals of BSCCO for studying XPS instead of the actual device as the minimum spot size of our X-ray beam is ~0.5~mm. Photoemission measurements were carried out using a Phoibos150 analyzer and monochromatic Al K$\alpha$ X-ray source at a base pressure of 8~$\times {10^{-11}}$ torr on an \emph{in-situ} cleaved BSCCO surface. We then investigate evolution of the core level electronic spectra of the composing elements by \emph{in-situ} deposition of Cr on freshly cleaved BSCCO surface. Different thicknesses of Cr (1~nm and 3~nm) were deposited by e-beam evaporation with a evaporation rate of 1~{\AA}/min, keeping the sample at room temperature.

\clearpage


\widetext
\begin{center}
\textbf{\large Supplementary Materials: On-demand local modification of high-$T_\text{c}$ superconductivity in few unit-cell thick Bi$_2$Sr$_2$CaCu$_2$O$_{8+\delta}$}
\end{center}

\setcounter{section}{0}
\setcounter{equation}{0}
\setcounter{figure}{0}
\setcounter{table}{0}
\setcounter{page}{1}

\renewcommand{\thesection}{S\arabic{section}}
\renewcommand{\theequation}{S\arabic{equation}}
\renewcommand{\thefigure}{S\arabic{figure}}
\renewcommand{\thepage}{S\arabic{page}}
\renewcommand{\bibnumfmt}[1]{[S#1]}
\renewcommand{\citenumfont}[1]{S#1}

 \makeatletter
\def\@fnsymbol#1{\ensuremath{\ifcase#1\or \dagger\or *\or \ddagger\or
   \mathsection\or \mathparagraph\or \|\or **\or \dagger\dagger
   \or \ddagger\ddagger \else\@ctrerr\fi}}
    \makeatother

\section{Device shaping}

\begin{figure}[h]
\includegraphics[width=15.5cm]{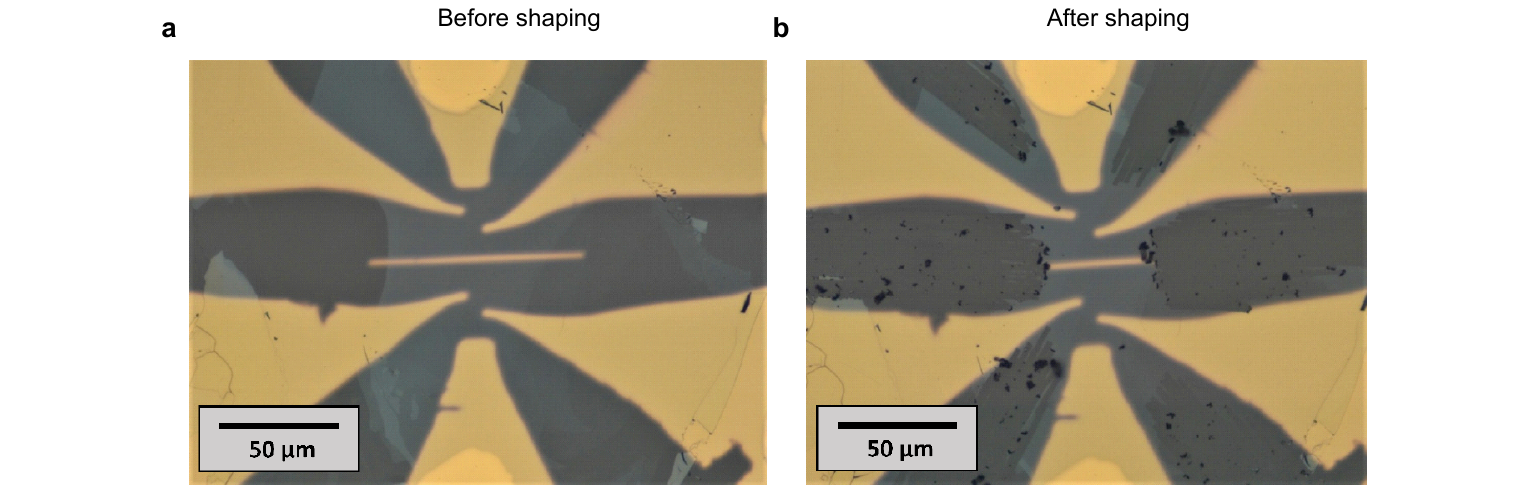}
\caption{ \label{fig:figS1}{\footnotesize Optical micrograph of BSCCO device before and after shaping. (a) Device image after making contacts to BSCCO flake with Au electrodes and a pattern of Cr line deposited in the middle of the device by shadow mask evaporation. (b) Device after removing unwanted parts of BSCCO flake with a sharp tip.}}
\end{figure}

We made all our BSCCO devices from thin flakes of BSCCO, exfoliated by anodic bonding \cite{balan_anodic_2010,sterpetti_comprehensive_2017} method. We select thin flakes of BSCCO for device fabrication by looking into optical contrast. We then do three successive evaporations through three different shadow masks \citep{deshmukh_nanofabrication_1999} for making Au electrodes (70~nm) to BSCCO, for making desired pattern (15~nm Cr or Ti) on BSCCO and extending the Au contacts (5~nm Cr and 70~nm Au) to BSCCO to larger pads respectively. The last set of Cr/Au electrodes are for wire bonding purpose and is in direct contact to Au electrodes only. Fig.~\ref{fig:figS1}a shows an optical image of the device after all the steps of evaporation. As seen from Fig.~\ref{fig:figS1}a, there are unwanted parts of BSCCO flake which are not covered with Cr line pattern. In transport measurement these unwanted parts of BSCCO will bypass the effect of Cr patterning. To avoid that, we shape the flake by cutting those undesired parts with a sharp tip. Fig.~\ref{fig:figS1}b shows optical image of the final device after shaping. The entire processes of exfoliation and mask alignments take place in the ambient environment. We try to minimize this exposure time to ambient, with typical exposure time $\sim$ 1 hr 30 min, and immediately cool down the device in a closed cycle cryostat. The measurements are done at high vacuum ($\sim$ 1~$\times {10^{-6}}$ mbar) using low-frequency ac lock-in detection technique in four-probe configuration.

\section{Devices with different widths of deposited \texorpdfstring{C\MakeLowercase{r}}{Cr} line pattern}

\begin{figure}[h]
\includegraphics[width=15.5cm]{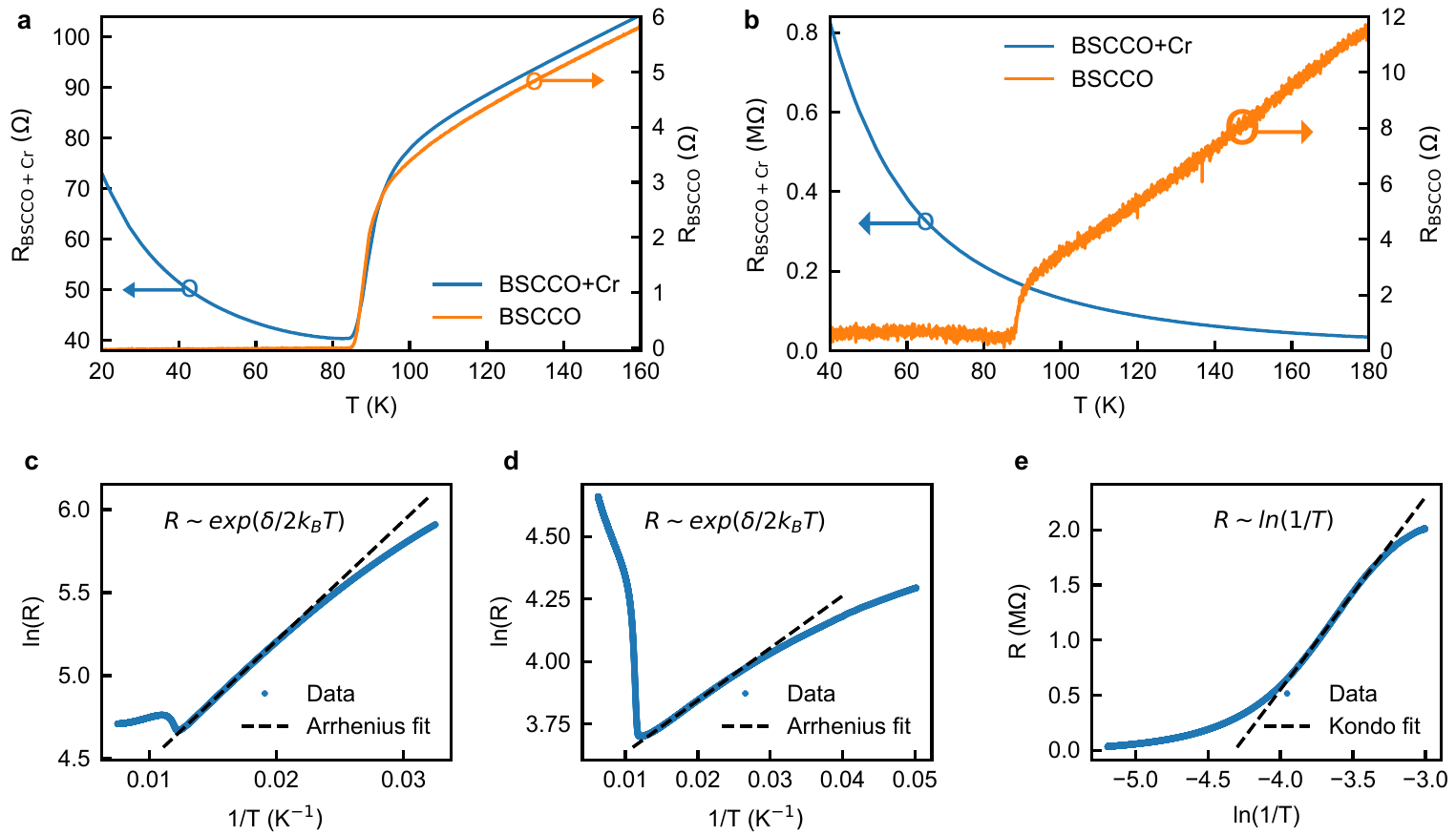}
\caption{ \label{fig:figS2} {\footnotesize Calculation of barrier energy for different widths of the deposited Cr line pattern. (a) $R$ vs. $T$ curve of a BSCCO device with \textbf{\SI{1.5}{\micro\metre}} wide Cr line. (b) $R$ vs. $T$ curve of BSCCO device with \SI{4}{\micro\metre} wide Cr line. As seen from the data, region across Cr pattern in this case is highly insulating even much above $T_\text{c}$. (c) Arrhenius fit of the resistance across Cr line below $T_\text{c}$ of the data presented in the main text. The fit (black dashed line) to the data is only limited to a small range of temperature. We estimate the energy gap ($\delta$=12.44~meV) from the segment of the data that agrees well with the Arrhenius model. (d) Arrhenius fit of the data across Cr pattern (Fig.~\ref{fig:figS2}a) below $T_\text{c}$. It also shows deviation from Arrhenius fit at lower temperatures. (e) Fitting of resistance across \SI{4}{\micro\metre} wide Cr patterned device (Fig.~\ref{fig:figS2}b) with Kondo model.}}
\end{figure}

We made two more class of devices with varying width of Cr pattern. Fig.~\ref{fig:figS2}a shows $R$ vs. $T$ curve for \SI{1.5}{\micro\metre} wide Cr line patterned on BSCCO. Normal state resistance across Cr pattern has metallic $\frac{dR}{dT}>0$, while below $T_\text{c}$ it has insulating $\frac{dR}{dT}<0$ response. We fit the insulating response below $T_\text{c}$ with Arrhenius model (Fig.~\ref{fig:figS2}d). Fig.~\ref{fig:figS2}b shows $R$ vs. $T$ curve for \SI{4}{\micro\metre} wide Cr patterned device. The region across Cr pattern has become highly resistive. Moreover, in this device we see $\frac{dR}{dT}<0$ in entire temperature range. In this case the Arrhenius model does not fit the data well. Assuming magnetism of Cr can play role in scattering, we fit the data with Kondo model as shown in Fig.~\ref{fig:figS2}e. Fig.~\ref{fig:figS2}c shows Arrhenius fit of the resistance across Cr patterned device presented in the main text. Because of spreading the width of the Cr pattern became \SI{1.8}{\micro\metre}. The energy gap estimated for this device is $\delta$=12.44~meV. Therefore, we find that the insulating barrier formed in BSCCO underneath Cr pattern increases with increasing width of the Cr pattern. Although the data fits quite well with Arrhenius model it deviates at lower temperatures. Moreover, to check whether the insulating nature is associated with the Kondo physics due to magnetism of Cr, further studies are needed.

\section{XPS study}

\begin{figure}[h]
\includegraphics[width=8cm]{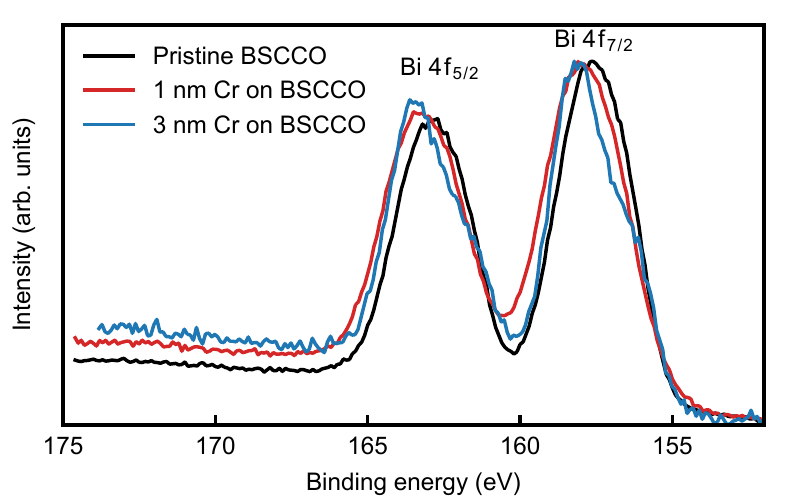}
\caption{ \label{fig:figS3}{\footnotesize Evolution of core-level electronic spectra of Bi 4\textit{f}. We see that with the deposition of Cr, additional peak at higher binding energy appears for both Bi 4\textit{f}$_{5/2}$ and Bi 4\textit{f}$_{7/2}$ along with the low intensity original peaks.}}
\end{figure}


Fig.~\ref{fig:figS3} shows Bi 4\textit{f} spectra of pristine BSCCO along with Cr covered BSCCO. For pristine BSCCO, the spectra exhibits two intense features corresponding to spin-orbit split Bi 4\textit{f}$_{7/2}$ and Bi 4\textit{f}$_{5/2}$ signal. With the deposition of 1~nm Cr layer on top, both the peaks shift to higher binding energy due to reduction in hole concentration \citep{harima_chemical_2003}. With the deposition of 3~nm Cr on top of BSCCO, each of the spin-orbit split Bi 4\textit{f} signal develops two distinct peaks. The intense feature at about 158.2~eV appears due to Bi at the interface of Cr and BSCCO and the weaker feature at 156.5~eV is the bulk contribution. Clearly, the Bi at the interface are more oxidized than the bulk Bi. XPS data for O and Cu presented in the main text along with the data for Bi presented here clearly suggest that deposition of Cr layer on the surface of BSCCO depletes the hole concentration in BSCCO underneath.

\section{More device applications}

\begin{figure}[h]
\includegraphics[width=15.5cm]{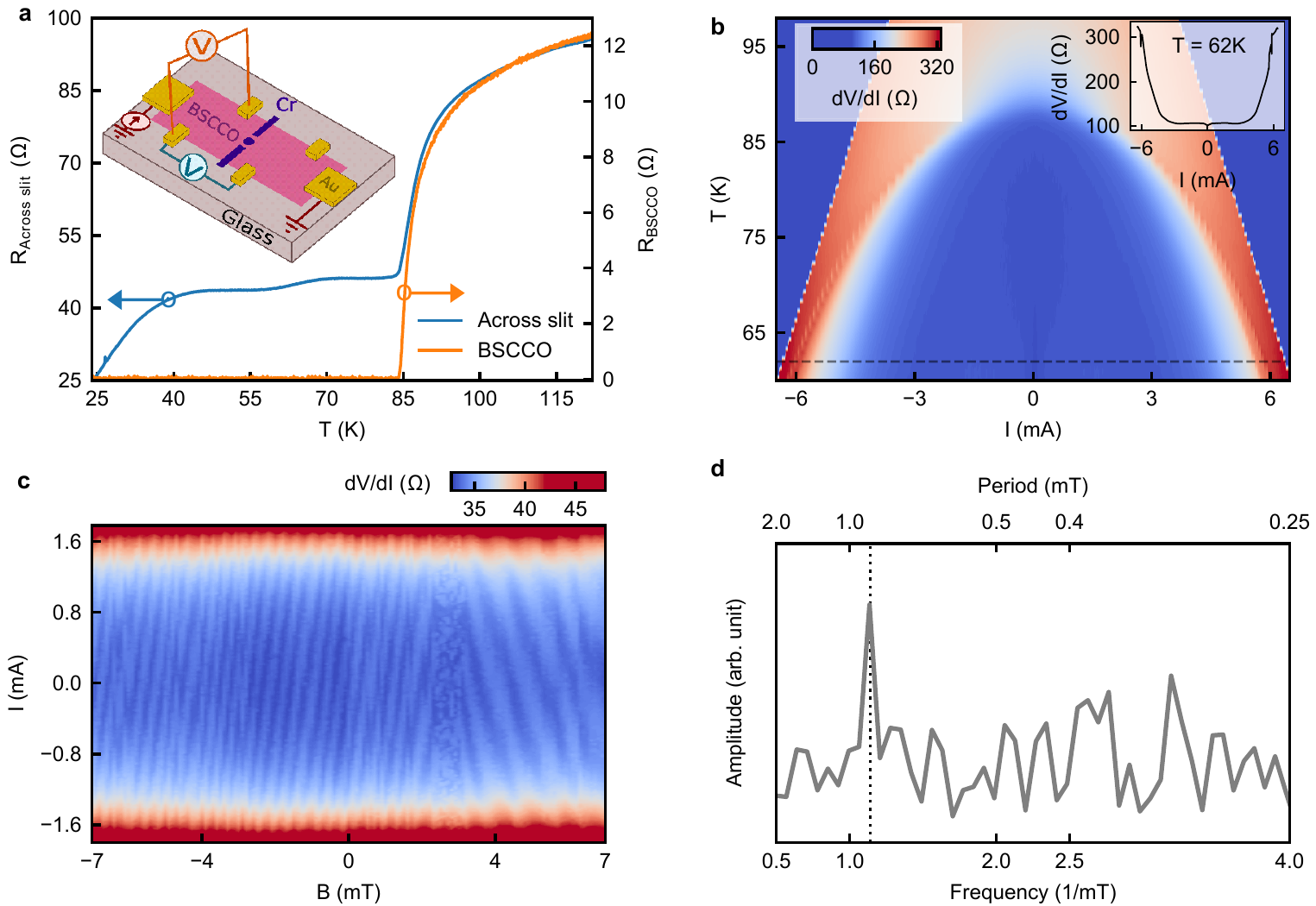}
\caption{ \label{fig:figS4} Realizing superconducting two-slit device with \textit{etch-a-sketch} patterning of superconductivity in BSCCO (a) $R$ vs. $T$ curve of patterned device with two slits. Inset shows the device schematic. The dimensions of the slits are -- width $\sim$ \SI{1}{\micro\metre}, gap \SI{1}{\micro\metre}. Resistance across Cr double slit has a drop at $T_\text{c}$ of BSCCO and has a downward trend at lower temperature. (b) Color scale plot of differential resistance $\frac{dV}{dI}$ across Cr slit as a function of dc biasing current ($I$) and temperature ($T$). Inset shows $\frac{dV}{dI}$ vs. $I$ at 62~K as a line cut of the color scale plot. The blue shaded area at the corners of the color scale plot is devoid of data. At higher temperatures we restricts the current to low values to avoid heating related damages to the device. (c) Color scale plot of $\frac{dV}{dI}$  vs. $I$ across Cr slits as a function of applied magnetic field ($B$) perpendicular to the sample plane at 85~K. We see the critical current oscillations with externally applied $B$. (d) FFT of the data shown in Fig.~\ref{fig:figS4}c. We see a peak in FFT amplitude at a frequency of $\approx$ 1.14~(mT)$^{-1}$ which corresponds to the period of oscillation being $\approx$ \SI{877}{\micro\tesla}. }
\end{figure}

Along with the patterned narrow BSCCO wire presented in the main text, here we also demonstrate a device with two narrow slits like geometry patterned on BSCCO. To realize it, we deposit a \SI{1}{\micro\metre} wide Cr line on few unit cell thick BSCCO having two slits of \SI{1}{\micro\metre} width separated by a circular area of $\sim$ \SI{1}{\micro\metre} diameter, as shown schematically in the inset of Fig.~\ref{fig:figS4}a. The two slits effectively define narrow constrictions of superconductor by making Cr covered BSCCO insulating; this leads to a superconducting loop of BSCCO. The actual width and length of the slits are limited by the spatial spreading of the pattern during shadow mask evaporation. The $R$ vs. $T$ behavior of this device geometry is shown in Fig.~\ref{fig:figS4}a. The resistance across double-slit pattern drops to $\approx$ 45~$\Omega$ at $T_\text{c}$ of pristine BSCCO and has a downward trend at temperatures below $T_\text{c}$ which is in contrast to the response of continuous Cr line pattern presented in the main text suggesting proximity effects. However, the resistance does not go to zero in our accessible range of temperature. Fig.~\ref{fig:figS4}b shows differential resistance $\frac{dV}{dI}$ across double-slit pattern as a function of dc biasing current ($I$) and temperature ($T$). A line cut at 62~K is shown in the inset of the figure showing $\frac{dV}{dI}$ vs. $I$. Unlike the continuous Cr line pattern, here we see a dip in differential resistance at zero bias current. Fig.~\ref{fig:figS4}c shows $\frac{dV}{dI}$ across double-slit as a function of biasing current and perpendicular magnetic field ($B$) at 85~K where we see the modulation of critical current $I_\text{c}$ with $B$. Fast Fourier Transform (FFT) of the data (Fig.~\ref{fig:figS4}c) is shown in Fig.~\ref{fig:figS4}d. The peak in FFT amplitude at frequency of 1.14~(mT)$^{-1}$ corresponds to a period of oscillation of $I_\text{c}$ with $B$ to be $\approx$ \SI{877}{\micro\tesla}. The dimension of the area of flux modulation extracted from the period of $B$ is $\approx$ 2.2~($\mu$m$)^{2}$ and this agrees well with the Cr covered BSCCO area separating two slits taking into account the spreading in mask evaporation. This suggests that there is some coherent transport across the two slits close to $T_\text{c}$. As the coherence length (typically $\sim$ 1-2~nm) diverges as $\sim \frac{1}{\sqrt{(1-T/T_\text{c})}}$, we intentionally use a temperature close to $T_\text{c}$ for this demonstration. The yield of fabricating these kind of superconducting interference devices is small (because of small coherence length). This can be improved by reducing minimum feature size of the patterns by \textit{h}-BN shadow mask.

\section{Reduction of critical current ($I_\text{c}$) of BSCCO wire}

\begin{figure}[h]
\includegraphics[width=8cm]{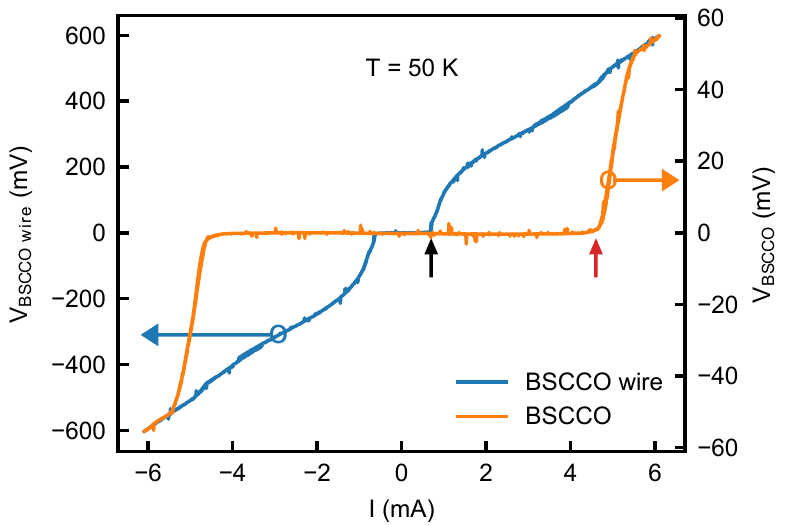}
\caption{ \label{fig:figS5}{\footnotesize Critical current ($I_\text{c}$) of BSCCO wire. We see that with the patterning of BSCCO in a wire geometry, $I_\text{c}$ reduces from that of pristine BSCCO. At 50~K $I_\text{c}$ of BSCCO is 4.6~mA (marked by red arrow) while it reduces to 0.7~mA across patterned BSCCO wire (marked by black arrow).}}
\end{figure}

Fig.~\ref{fig:figS5} shows dc current-voltage characteristic ($I$-$V$) at 50~K of the BSCCO wire geometry device presented in the main text. The patterning of BSCCO in wire geometry reduced the $T_\text{c}$ across wire. Here we see that across patterned BSCCO wire $I_\text{c}$ also reduced from that of pristine BSCCO. Black and red arrow mark the $I_\text{c}$ across BSCCO wire (0.7~mA) and pristine BSCCO (4.6~mA) at 50~K as shown in Fig.~\ref{fig:figS5}.


\end{document}